\documentclass[aps, prd, twocolumn, showpacs, superscriptaddress, nofootinbib,floatfix]{revtex4-2}
\usepackage{amsmath,mathrsfs,amsbsy,color,graphicx,bm,amsthm,amsfonts}
\usepackage{bbm}
\usepackage{bm}
\usepackage{times}
\usepackage{units}
\usepackage{dcolumn}
\usepackage{graphicx}
\usepackage{epsfig}
\usepackage{epstopdf}
\usepackage[colorlinks,linkcolor=red,anchorcolor=green,citecolor=blue,CJKbookmarks=True]{hyperref}
\DeclareMathSymbol{\shortminus}{\mathbin}{AMSa}{"39}
\usepackage{mathrsfs}
\usepackage{braket}
\usepackage{amssymb}
\usepackage{txfonts}
\usepackage{enumitem}
\usepackage{multirow}
\graphicspath{{fihures/},{pics/}} \usepackage{subfigure} 
\usepackage{microtype}
\usepackage[cmyk]{xcolor} 
\usepackage{placeins}

\usepackage{orcidlink}

\begin{document}
\title{Wormhole-Induced correlation: A Link Between Two Universes}

\author{Zhilong Liu~\orcidlink{0009-0000-0353-5113}}
\affiliation{Department of Physics, Key Laboratory of Low Dimensional Quantum Structures and Quantum Control of Ministry of Education, Hunan Research Center of the Basic Discipline for Quantum Effects and Quantum Technologies,  and Synergetic Innovation Center for Quantum Effects and Applications, Hunan Normal
	University, Changsha, Hunan 410081, P. R. China}

\author{Wentao Liu~\orcidlink{0009-0008-9257-8155}}
\affiliation{Department of Physics, Key Laboratory of Low Dimensional Quantum Structures and Quantum Control of Ministry of Education, Hunan Research Center of the Basic Discipline for Quantum Effects and Quantum Technologies,  and Synergetic Innovation Center for Quantum Effects and Applications, Hunan Normal
	University, Changsha, Hunan 410081, P. R. China}

\author{Xiaofang Liu~\orcidlink{0000-0002-9450-4001}}
\affiliation{Department of Physics, Key Laboratory of Low Dimensional Quantum Structures and Quantum Control of Ministry of Education, Hunan Research Center of the Basic Discipline for Quantum Effects and Quantum Technologies,  and Synergetic Innovation Center for Quantum Effects and Applications, Hunan Normal
	University, Changsha, Hunan 410081, P. R. China}

\author{Jieci Wang~\orcidlink{0000-0001-5072-3096}}
\email{jcwang@hunnu.edu.cn (Corresponding author)}\affiliation{Department of Physics, Key Laboratory of Low Dimensional Quantum Structures and Quantum Control of Ministry of Education, Hunan Research Center of the Basic Discipline for Quantum Effects and Quantum Technologies,  and Synergetic Innovation Center for Quantum Effects and Applications, Hunan Normal
	University, Changsha, Hunan 410081, P. R. China}

\begin{abstract}
Motivated by the profound connection between quantum mechanics and spacetime geometry, particularly the conjectured correspondence between wormholes and quantum entanglement as proposed in the ER=EPR framework, this study investigate the influence of wormhole geometries on quantum information extraction. We examine the correlation-specifically mutual information (MI) and entanglement-extracted by two Unruh-DeWitt (UDW) detectors from the quantum vacuum field in the presence of a Bañados-Teitelboim-Zanelli (BTZ) wormhole featuring a null-like throat, also known as an Einstein-Rosen bridge. First, we analyze how the detector's position relative to the wormhole throat and the throat's size affect the extracted MI. Our results indicate that the wormhole enhances MI extraction, with maximal MI achieved when the detectors are located at specific image-symmetric points connected by the wormhole. By analyzing the behavior of the nonlocal contribution term and the classical noise term, it is found that the correlations extracted contain genuine non-classical components. This work highlights the feasibility of extracting quantum correlations through null-like wormhole geometries and provides a novel perspective for probing the potential relationship between spacetime topology and the nonlocal characteristics of quantum mechanics.
\end{abstract}
\pacs{~}

\maketitle
\noindent
\section{introduction}

The nonlocality of quantum theory has long been a central and intensely debated topic, first brought to prominence by Einstein and his collaborators in what is now famously known as the Einstein-Podolsky-Rosen (EPR) paradox \cite{Einstein:1935rr}. Although their original aim was to question the completeness of quantum mechanics by suggesting that hidden variables might account for the apparent nonlocal behavior, this paradox ultimately highlighted a fundamental feature of quantum theory. This distinctive characteristic of quantum systems, which contradicts classical notions of locality and realism, has since been experimentally verified and widely recognized \cite{Bell:1964kc,Freedman:1972zza,Aspect:1981nv,Bouwmeester:1997slj}. Furthermore, it has demonstrated that nonlocal correlations arising from vacuum fluctuations can lead the vacuum state in quantum field theory to maximally violate Bell inequalities \cite{Summers:1985pzz,Valentini:1991eah,Reznik:2003mnx}.
 	Recently,  subsequent studies have shown that nonlocal correlations can be extracted  from the  vacuum  \cite{Reznik:2002fz,Unruh:1976db,Salton:2014jaa}, a process commonly referred to as correlation harvesting. Such protocol has been extended to curved spacetime scenarios, which has become an important branch in the field of relativistic quantum information \cite{Fuentes-Schuller:2004iaz,Ahn:2008zf,Zhou:2022nur,Zhou:2021nyv,Wu:2022lmc,Wu:2022xwy,Liu:2023zro,Chen:2023xbc,Wu:2023sye,Wu:2023spa,Wu:2024pwa,Liu:2024pse,Liu:2024wpa,Liu:2024yrf,Ji:2024fcq,AraujoFilho:2024ctw,AraujoFilho:2025hkm}.
	It was demonstrates that correlation harvesting is sensitive to influences such as detector motion, switching dynamics, boundary conditions, spacetime structure, topology, and even indefinite causal orders \cite{Henderson:2017yuv,Bueley:2022ple,Zhang:2020xvo,Gallock-Yoshimura:2021yok,Pozas-Kerstjens:2015gta,Cong:2018vqx,Cong:2020nec,Svidzinsky:2024tjo,VerSteeg:2007xs,Kukita:2017etu,Ng:2018drz,Wu:2024qhd,Xu:2020pbj,Tjoa:2020eqh,Martin-Martinez:2015qwa,Foo:2020dzt,Liu:2025bpp,Foo:2020xqn}.
	
	
	On the other hand, in 1935, Einstein and Rosen introduced the concept of a bridge, later termed the ``Einstein-Rosen (ER) bridge," to describe a structure connecting two congruent sheets of spacetime \cite{Einstein:1935tc}. 
	This construct, commonly referred to as a wormhole, was initially thought to be untraversable \cite{Fuller:1962zza,Friedman:1993ty,Galloway:1999bp}. 
	Nevertheless, it has attracted substantial interest in the study of spacetime geometry and its broader implications \cite{Saad:2021rcu,Forste:2024nsw,Hollowood:2024uuf}.
	Recently, studies have suggested that the ER bridge connecting two black holes is generated by nonlocal correlations between the microstates of the two black holes, a conjecture now widely recognized as the ER=EPR relation \cite{Maldacena:2001kr,Maldacena:2013xja}. 
	This profound idea has stimulated extensive investigations into the relationship between wormhole structures and quantum theory \cite{Jusufi:2023dix,Kain:2023ore,Dai:2020ffw}, offering new insights into the interplay between spacetime and quantum mechanics. Such nonlocality, originating from the spacetime structure, can influence quantum fields within this background. Consequently, it becomes manifest in the vacuum fluctuations of the quantum field. The entanglement carried by these vacuum fluctuations can then be extracted using the UDW detector model.
	
	To explore the interplay between wormhole geometry and quantum nonlocality, this study investigates the harvesting of quantum correlations in a spacetime containing a Bañados-Teitelboim-Zanelli (BTZ) wormhole with a null-like throat. Specifically, we employ a pair of UDW particle detectors to analyze how the throat size of the wormhole and the spatial arrangement of the detectors affect the extracted correlations. Our results show that the wormhole structure acts as a geometric bridge enabling the emergence of quantum correlations between classically disconnected regions. The harvested correlations exhibit both local and nonlocal components, with the both enhanced as the detectors approach symmetric positions relative to the throat. Notably, when the throat size approaches zero—effectively in the absence of a wormhole—these correlations vanish, underscoring their geometric origin. It should be underscored that these correlation harvesting phenomena do not imply any possibility of communication between the two universes, as classical signals cannot traverse the wormhole.
	
	The structure of this paper is organized as follows. In Section~\ref{sec_II}, we introduce the concept of the BTZ wormhole, present its spacetime embedding diagram, and discuss its relation to the Penrose diagram. We then briefly review the framework of the UDW detector model. Section~\ref{sec_III} presents our analysis and numerical results regarding the harvested correlation, including both MI and concurrence. Finally, Section~\ref{sec_IV} provides a comprehensive summary of our findings and conclusions, with additional details of the numerical calculations provided in the Appendix~\ref{sec_V}.
\section{The Model} \label{sec_II}
\subsection{BTZ Wormhole}
	The BTZ black hole is a solution that exists within a (2+1)-dimensional Anti-de Sitter (AdS) spacetime \cite{Banados:1992gq}. 
	Due to its low-dimensional nature and absence of physical singularities,  it serves as an ideal theoretical framework for investigating the AdS/CFT correspondence and various aspects of quantum gravity theories \cite{Horowitz:1993jc,Deger:2022qob,chen2024adscftcorrespondencehyperboliclattices,Liu:2022dcn}. 
	Additionally, this makes it a valuable tool in advancing research in related quantum theories. 
	The spacetime metric of the BTZ black hole in static coordinates is given by
	\begin{equation}
		\begin{aligned}
			&\mathrm{d}s^2=-f(r)\mathrm{d}t^2+\frac{1}{f(r)}\mathrm{d}r^2 + r^2\mathrm{d}\phi^2, \\
			&f(r)=\frac{r^2}{l^2}-M,
		\end{aligned}
	\end{equation}
	where $M$ is the mass of BTZ black hole and $l$ is related to the negetive cosmic constant $\Lambda$ with the relation $l^2=-\frac{1}{\Lambda}$. The BTZ black hole has event horizon at $r_h=l\sqrt{M}$. 
	
\begin{figure}[htbp]
		\centering{\includegraphics[width=0.98\linewidth]{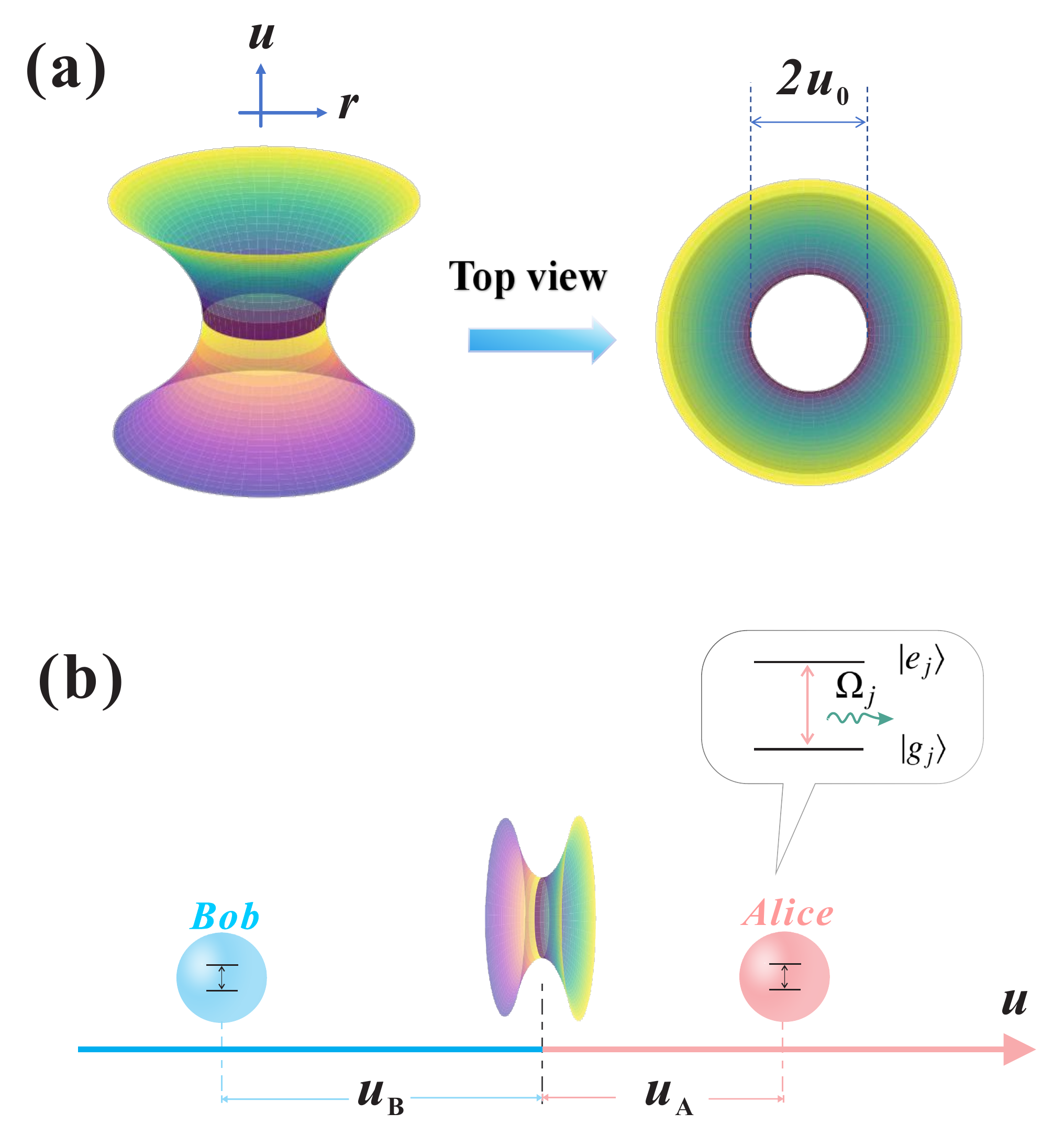}
		}
		\caption{ (a) Spacetime embedding graph of the BTZ null-like wormhole, depicted at a constant time slice $t$ = \text{const}. The upper and lower regions represent two congruent universes connected by the central circle, which represents the throat of the wormhole. The wormhole vanishes when the radius $u_0$ approaches zero. (b) Schematic representation of the throat size and the arrangement of UDW detectors during our correlation harvesting process. Detectors Alice and Bob are positioned on opposite sides of the wormhole, initially in their ground states, uncorrelated, and possessing the same energy gap \( \Omega_A=\Omega_B=\Omega \).
		} 
		\label{FIG_1}
	\end{figure}

	We can extend the region into a null-like wormhole spacetime through the following coordinate transformation \cite{Einstein:1935tc,Anand:2024vmy} 
	\begin{equation}
		u^2=r-r_h,
	\end{equation}
	and then the null-like wormhole spacetime metric can be written as
	\begin{equation}
		\mathrm{d}s^2=-\frac{u^2(u^2+2u_0)}{l^2}\mathrm{d}t^2+\frac{4l^2}{u^2+2u_0}\mathrm{d}u^2 + (u^2+u_0)^2\mathrm{d}\phi^2,           
	\end{equation}
	where $u_0=r_h$ represents the radius of the wormhole throat, and the wormhole ceases to exist when $u_0=0$. The coordinate $u \in (-\infty,+\infty)$ spans the entire wormhole structure, with the positive and negative signs of $u$ corresponding to two congruent universes, as illustrated in Fig. \ref{FIG_1}(a). The radial coordinate $r$ varies from $+\infty$ to $r_h$ and then from $r_h$ back to $+\infty$ as $u$ transitions from $-\infty$ to $+\infty$. 
	The connecting channel of the two universes is called the Einstein-Rosen bridge, or wormhole, with a null throat at $u=0$. Setting $\mathrm{d}\phi=0$, the speed of radial null geodesic at the throat is found to be zero,
	\begin{equation}
		\frac{du}{dt}|_{u=0}=\pm\frac{u(u^2+2u_0)}{2l^2}|_{u=0}=0.
	\end{equation}

\begin{figure}[t]
		\centering{\includegraphics[width=0.56\linewidth]{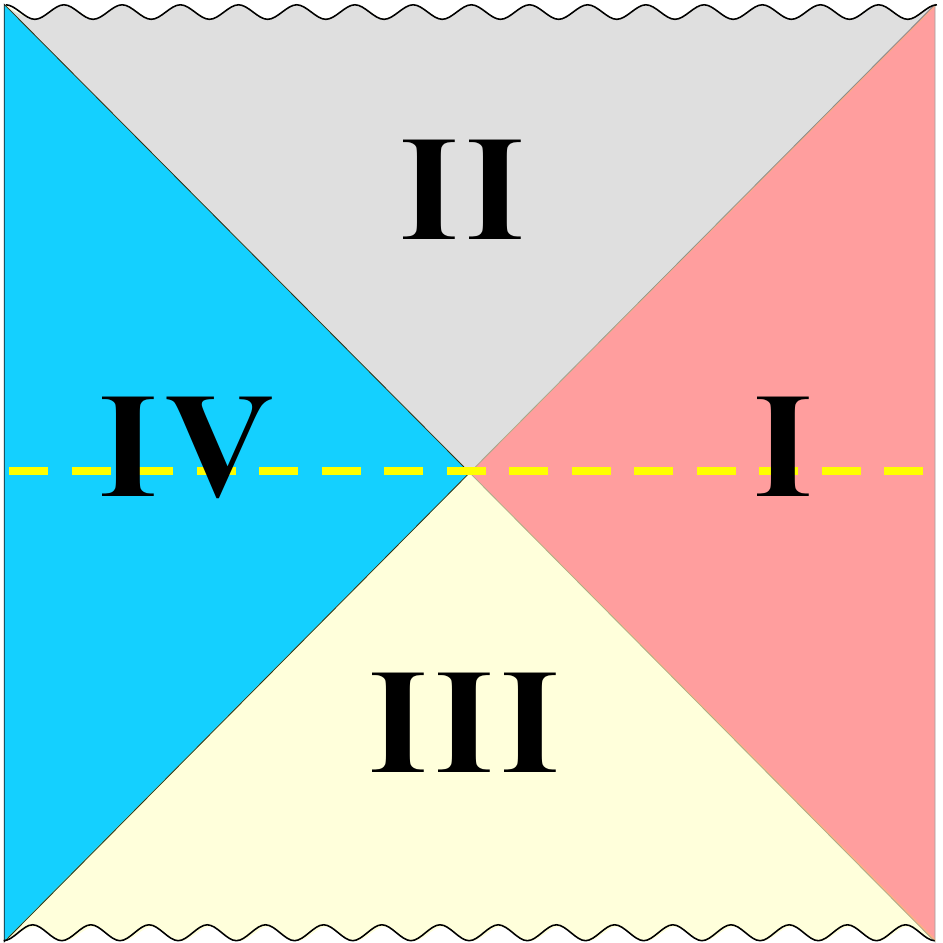}
		}
		\caption{The Penrose diagram of the extended BTZ spacetime. Regions I and IV as two universes connected via an event horizon. Regions II and III represent the interiors of the black hole and white hole, respectively. The wavy lines at the top and bottom denote the spacelike infinity ($r=0$), while special spacelike hypersurfaces ($U = -V$ or $T = 0$) traversing both universes are indicated by yellow dashed lines.} 
		\label{Penrose}
	\end{figure}

	To gain a deeper and more intuitive understanding of the BTZ wormhole spacetime, we briefly introduce its Penrose diagram and provide a visual interpretation within this framework. We adopt the Eddington–Finkelstein coordinates, defined as
	\begin{equation}
		\tilde{v} = t + r_*, \quad \tilde{u} = t - r_*,    
	\end{equation}
	where $r_*$ denotes the tortoise coordinate and is given by
	\begin{align}
		r_* &= \int \frac{1}{f(r)} \mathrm{d}r =l^2\int \left[ \frac{1}{2r_h(r-r_h)}-\frac{1}{2r_h(r+r_h)} \right] \mathrm{d}r, \nonumber \\
		&=\frac{l^2}{2r_h}ln\left|\frac{r-r_h}{r+r_h}\right|.
	\end{align}
	We then employ Kruskal coordinates, defined as
	\begin{align}
		&T=\frac{U+V}{2}, \quad R=\frac{V-U}{2}, \nonumber \\
		&U = \pm\frac{e^{-\kappa \tilde{u}}}{\kappa}, \quad V = \pm \frac{e^{\kappa \tilde{v}}}{\kappa},
	\end{align}
	where $\kappa=\sqrt{M}/l$ represents the surface gravity of the black hole.  Using these coordinates, we construct the Penrose diagram for the BTZ spacetime, as shown in Fig. \ref{Penrose}. Regions I and IV correspond to two distinct universes with \( UV < 0 \), while regions II and III represent the black hole and white hole with \( UV > 0 \), respectively. These coordinates satisfy the following relations:  
	\begin{align}
		T^2-R^2&=UV= \pm 1/\kappa^2 e^{2\kappa r_*} , \nonumber\\
		&=\pm g(r)|r-r_h| =-g(r)(r-r_h),\\
		g(r)&=\frac{1}{\kappa^2(r+r_h)}e^{\frac{\kappa l^2}{r_h}}.
	\end{align}
	Setting the coordinate \( T = 0 \) yields  
	\begin{equation}
		R=\pm \sqrt{g(r)(r-rh)},
	\end{equation}
	which corresponds to the yellow dashed line in Fig~\ref{Penrose}, representing the two universes (regions where \( R > 0 \) and \( R < 0 \)) connected by a wormhole with a throat at \( r = r_h \). This configuration closely resembles the wormhole geometry described in our model, particularly in the case where \( g(r) \) remains constant.
	
	In fact, unlike traversable wormholes \cite{Morris:1988tu,Morris:1988cz,Lobo:2020kxn}, the throat of this BTZ wormhole corresponds to the event horizon in the original black hole spacetime. This implies that classical particles are unable to traverse the throat. As a result, this type of wormhole is referred to as an ``untraversable wormhole".  In this paper, we will investigate whether this classically untraversable wormhole can induce quantum correlations between the two congruent universes.
	
	Due to the BTZ can be constructed by a topological image identification from $AdS_3$ spacetime, i.e $\Gamma$:$(t,r,\phi)$ $\rightarrow$ $(t,r,\phi+2\pi)$ \cite{Avis:1977yn,Lifschytz:1993eb}. Considering a conformally coupled scalar field $\hat{\phi}(x)$ in the extended BTZ wormhole spacetime and the Hartle-Hawking vacuum $\ket{0}$. The Wightman function for this wormhole spacetime, denoted as $W_{wh}(\mathrm{x,x^\prime})$ := $\bra{0}\hat{\phi}(\mathrm{x})\hat{\phi}(\mathrm{x^\prime})\ket{0}=\sum^{+\infty}_{n=-\infty}W_{AdS}(\mathrm{x,\Gamma^n x^\prime})$. Consequently, using the topological image identification method, the $W_{wh}$ can be written as 
	\begin{align}
		W_{wh}(\mathrm{x,x^\prime})=\frac{1}{4\pi\sqrt{2}l}\sum^{+\infty}_{n=-\infty}\left[\frac{1}{\sqrt{\sigma_\epsilon(\mathrm{x,\Gamma^n x^\prime})}}-\frac{\xi}{\sqrt{\sigma_\epsilon(\mathrm{x,\Gamma^n x^\prime})+2}}\right],
	\end{align}
	where
	\begin{align}
		\sigma_\epsilon(\mathrm{x,\Gamma^n x^\prime})&=U(u,u^\prime,u_0)\,\mathrm{cosh}\left[\frac{u_0}{l}(\Delta\phi-2\pi n)\right] \nonumber \\
		&-1 -\sqrt{u^2u^{\prime^2}}U(u,u^\prime,u_0)\,\mathrm{cosh}\left[\frac{u_0}{l^2}\Delta t-i\epsilon\right].
	\end{align}
	Here $U(u,u^\prime,u_0)=(u^2+2u_0)({u^\prime}^2+2u_0)/u_0^2$, $\Delta \phi:=\phi-\phi^\prime$, $\Delta t:=t-t^\prime$, and $\epsilon$ serve as an ultraviolet regulator. The constant nonumber $\xi \in \left\{-1,0,1\right\}$ represents different boundary condition at the spatial infinity: Neumann for $\xi=-1$, transparent for $\xi =0$ and Dirichlet for $\xi =1$. Throughout this paper, we will choose the Dirichlet boundary condition.
\subsection{UDW detecors}
	We consider two point-like detectors and each detector primarily consists of a two-level quantum system characterized by an energy gap $\Omega$ between its ground state and excited state. These detectors interact weakly with a conformally coupled scalar field $\hat{\phi}(\mathrm{x})$ through the following Hamiltonian
	\begin{equation}
		\hat{H}_j(\tau_j)=\mathrm{\lambda}_j \chi_j(\tau_j)\hat{\mu}_j(\tau_j)\otimes \hat{\phi}(\mathrm(x)_j(\tau_j)),
	\end{equation}
	where $j \in\left\{A,B\right\}$ specifies the different detectors, $\lambda_j$ is the weakly coupled coefficient, and $\chi_j$ represents switching function controlling the interaction duration of detector-$j$. In this study, we consider a Gaussian switching function with duration $\sigma$: $\chi_j(\tau_j)=e^{-\tau_j^2/2\sigma}$.The monopole moment $\hat{\mu}_j$ describes the dynamics of each detector and can be written as
	\begin{equation}
		\hat{\mu}_j=\ket{e_j}\bra{g_j}e^{i \Omega_j \tau_j}+\ket{g_j}\bra{e_j}e^{-i \Omega_j \tau_j},
	\end{equation}
	with $\ket{g_j}$, $\ket{e_j}$ are respectively represent the ground state and the excited state of each detector and $\Omega_j$ for corresponding energy gap. 
	Then the time evolution operator $\hat{U}_I$ can be given by a time ordering operator $\mathcal{T}$ as
	\begin{equation}
		\hat{U}_I= \mathcal{T}exp\left\{-i \int dt \left(\frac{d\tau_A}{dt}H_A +\frac{d\tau_B}{dt}H_B\right) \right\}.
	\end{equation}
	Due to the weakly interaction constant $\lambda$ is small, we can expand the time evolution operator $\hat{U}_I$ appling the Dyson series as $\hat{U}_I=\mathbbm{1} +\hat{U}^{(1)}_I +\hat{U}^{(2)}_I+ \mathcal{O}(\lambda^3)$.
	We then surppose the two detectors are both the same  (having the same coupled coefficient $\lambda$ and energy gap $\Omega$) and initially prepared in their ground state with initially uncorrelated scalar field being in the vacuum state $\ket{0}$. Therefore, the initial state dencity metrix can be written as
	\begin{equation}
		\rho_0=\ket{g_A}\bra{g_A}\otimes\ket{g_B}\bra{g_B}\otimes\ket{0}\bra{0}. 
	\end{equation}   
	Then the final state after a duration of interaction can be given by appling time evolution operator as $\rho_f =\hat{U}_I \rho_0 \hat{U}_I^\dagger $. Subsequently, we can get the reduced dencity metrix (RDM) $\rho_{AB}$ by tracing out the part of field and choosing the basis $\left\{\ket{g_A}\ket{g_B},\ket{g_A}\ket{e_B},\ket{e_A}\ket{g_B},\ket{e_A}\ket{e_B} \right\}$, then the RDM can be written as 
	\begin{equation}
		\rho_{AB}=
		\begin{bmatrix}
			1-\mathcal{L}_{AA}-\mathcal{L}_{BB} & 0 &0 & \mathcal{M}^*  \\
			0 & \mathcal{L}_{BB}  &\mathcal{L}_{AB}^* &0 \\
			0 & \mathcal{L}_{AB}  &\mathcal{L}_{AA} &0 \\
			\mathcal{M} & 0  &0 &0
		\end{bmatrix}+ \mathcal{O}(\lambda^4),
	\end{equation}
	where 
	\begin{align}
		\mathcal{L}_{ij}=&\lambda^2 \int_{\mathbb{R}}d\tau_i\int_{\mathbb{R}}d\tau_j^\prime \chi_i(\tau_i)\chi_i(\tau_j^\prime)e^{-i\Omega(\tau_i-\tau_j^\prime)} \nonumber \\
		&\times W_{wh}(\mathrm{x}_i(\tau_i),\mathrm{x}_j(\tau_j^\prime)), \\
		\mathcal{M}=&-\lambda^2 \int_{\mathbb{R}}d\tau_A\int_{\mathbb{R}}d\tau_B^\prime \chi_B(\tau_A)\chi_B(\tau_B^\prime)e^{i\Omega(\tau_A+\tau_B^\prime)} \nonumber \\
		&\times [\Theta\left[t(\tau_A)-t(\tau_B)\right]W_{wh}(\mathrm{x}_A(\tau_A),\mathrm{x}_B(\tau_B^\prime)) \nonumber \\
		&+\Theta\left[t(\tau_B)-t(\tau_A)\right]W_{wh}(\mathrm{x}_B(\tau_B),\mathrm{x}_A(\tau_A^\prime))].
	\end{align}
	\begin{figure}[t]
	\centering{\includegraphics[width=0.60\linewidth]{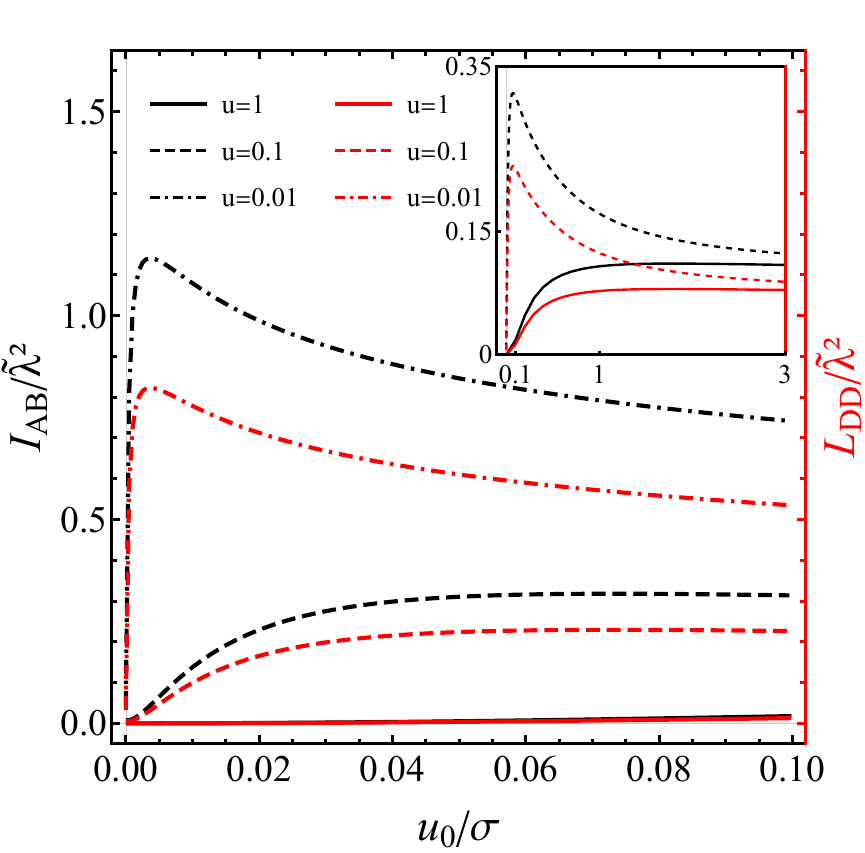}
		\caption{Plot of extracted MI $I_{AB}/\tilde{\lambda}^2$ (depicted in red) and TP $L_{DD}/\tilde{\lambda}^2$  (in blue) versus the size of the throat $u_0/\sigma$. Alice and Bob are symmetrically distributed around the throat at three different distances, represented by different line types: $u/\sigma=1$ (solid), $u/\sigma=0.1$ (dashed), and $u/\sigma=0.01$ (dotdashed). 
			The upper-right subfigure is plotted for the last two distributions with a larger value of  $u_0/\sigma$. Parameters are $\Omega \sigma = 1$, $M = 10^{-2}$.}
		\label{fig_rheffect}
		}
	\end{figure}
	In fact, by tracing out either Alice's or Bob's part of the reduced density matrix (RDM) $\rho_{AB}$,  we obtain the single-detector density matrix $\rho_{D},D\in\{A,B\}$ \cite{Martin-Martinez:2015qwa}:
	\begin{equation}
		\rho_{D}=
		\begin{bmatrix}
			1-\mathcal{L}_{DD} & 0   \\
			0 & \mathcal{L}_{DD} 
		\end{bmatrix}+ \mathcal{O}(\lambda^4).
	\end{equation}

	This resulting RDM $\rho_{D}$ indecates that the element $\mathcal{L}_{DD}$ correspond to the transition probabilities (TP) of each detector. Meanwhile, the off-diagonal element $\mathcal{M}$ and $\mathcal{L}_{AB}$ are associated with the quantum entanglement and MI shared between the UDW particle detectors.

\section{Result} \label{sec_III}
	Based on the UDW detector model, the interaction between two initially independent detectors and the vacuum field enables the extraction of correlations from the vacuum field to the two detectors. In this section, we primarily compute the matrix elements of the RDM $\rho_{AB}$ for the two detectors after the interaction: \( L_{DD} \) (which represents the TP), as well as \( L_{AB} \) and \( |\mathcal{M}| \) (which quantify the correlations between the two detectors). The MI $I_{AB}$  is then calculated as
	\begin{align}\label{LAABB}
		I_{AB}=&\mathcal{L}_{+}ln(\mathcal{L}_{+})+\mathcal{L}_{-}ln(\mathcal{L}_{-}) \nonumber \\
		&-\mathcal{L}_{AA}ln(\mathcal{L}_{AA})-\mathcal{L}_{BB}ln(\mathcal{L}_{BB})+\mathcal{O}(\lambda^4),
	\end{align}
	where
	\begin{equation}\label{LAB}
		\mathcal{L}_{\pm}:=\frac{1}{2}\left(\mathcal{L}_{AA}+\mathcal{L}_{BB}\pm \sqrt{(\mathcal{L}_{AA}-\mathcal{L}_{BB})^2+4\left|\mathcal{L}_{AB}\right|^2}\right).
	\end{equation}
	The MI \( I_{AB} \) is observed to exhibit a strong dependence on the element \( \mathcal{L}_{AB} \), such that when \( \mathcal{L}_{AB} \) vanishes, and the MI \( I_{AB} \) also becomes zero.
	Finally, we analyze the nonlocal component of these extracted correlations by computing the concurrence $C_{AB}$ \cite{Martin-Martinez:2015qwa,Smith:2017vle}
	\begin{equation}
		\label{eq:concurrence}
		C_{AB}=2 \, \mathrm{Max}\left\{0,|\mathcal{M}|-\xi\right\}+\mathcal{O}(\lambda^4),
	\end{equation}
	where
	\begin{equation}	\label{eq:noise}
		\xi=\sqrt{\mathcal{L}_{AA}\mathcal{L}_{BB}}.
	\end{equation}
	It is evident that the entanglement between the two detectors is nonzero when the contribution from the nonlocal term $|\mathcal{M}|$ exceeds that of the local noise term $\xi$~\cite{Gallock-Yoshimura:2021yok}.
	\begin{figure*}[htbp]
		\centering{\subfigure{\includegraphics[width=0.99\textwidth,height=0.33\textwidth]{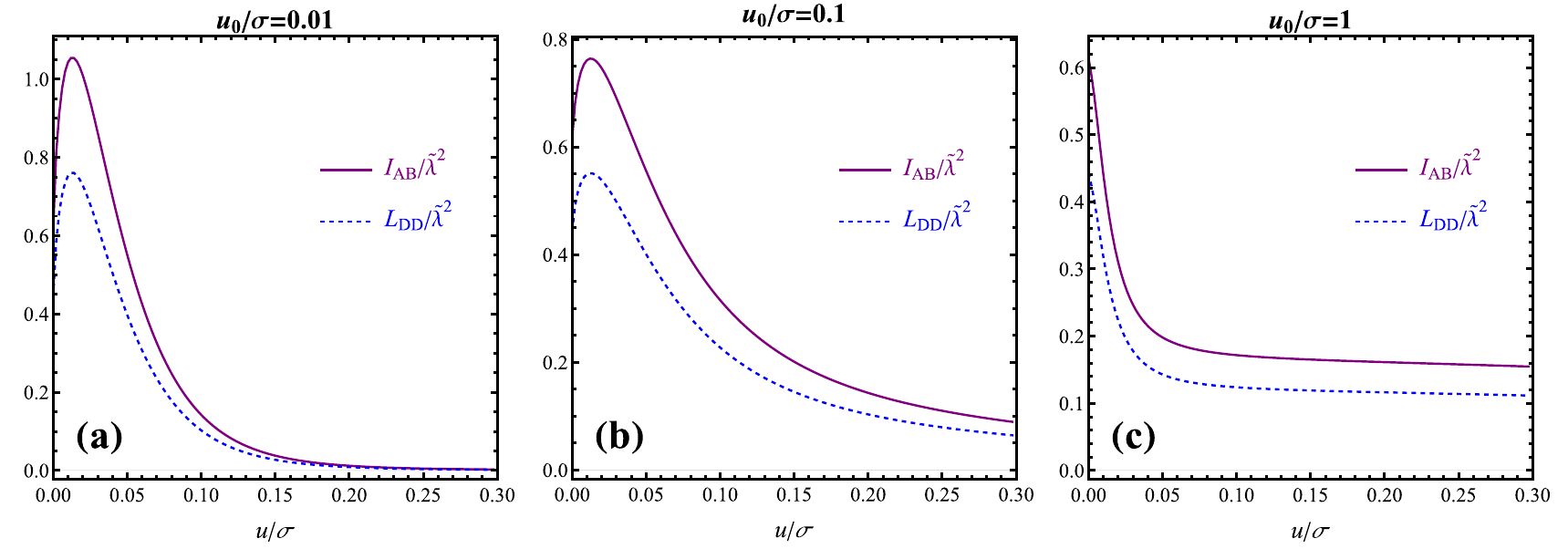}}
		}
		\caption{Plot of the extracted MI (solid purple) and TP (dashed blue). The particle detectors are symmetrically placed. From (a) to (c), the size of the wormhole throat is set to 0.01, 0.1, and 1, respectively. Here $\Omega \sigma = 1$, $M= 10^{-2}$.}
		\label{fig_schBi}
	\end{figure*} 	

	As illustrated in Fig~\ref{FIG_1} (b), the two detectors are positioned on either side of the wormhole throat  throughout the entire interaction duration. Given the congruence of the two universes, the TP satisfy \( L_{AA} = L_{BB} \) when the detectors are placed symmetrically along the \( u \)-axis, with the off-diagonal element \( \mathcal{L}_{AB} = \mathcal{L}_{DD} \). Consequently, in these configurations, the extracted correlations depend solely on the TP. To focus on the influence of the wormhole spacetime on the harvested detector-detector correlations, and for simplicity, we first investigate the variation of the TP and MI with the wormhole throat size under the condition of symmetric placement of the two detectors relative to the wormhole throat along the \( u \)-axis, i.e., \( u_A = -u_B = u \). The results are depicted by the blue and red lines, respectively, in Fig~\ref{fig_rheffect}, where we consider three different symmetric placements: \( u = 0.01 \) (dot-dashed line), \( u = 0.1 \) (dashed line) and \( u = 1 \) (solid line). 
	
	In these plots, both the extracted MI and TP exhibit consistent trends. This arises because, under the symmetric placement condition, the MI reduces to 
	\begin{equation}
		\label{eq:mi_symmetric}
		I_{AB} = 2ln2 \,\mathcal{L}_{DD}.
	\end{equation}
	Therefore, the MI $I_{AB}$ is proportional to the TP \( \mathcal{L}_{DD} \). These plot lines reveal a rapid increase to a peak as the wormhole throat size begins to expand, followed by a gradual decline to a relatively stable value. The peak heights increase as the two particle detectors are positioned closer together. When the wormhole throat is small, the detector distribution significantly affects the results. As the throat size increases, the extracted MI stabilizes across various distributions. Furthermore, we find that the extracted MI remains nonzero until the wormhole throat size approaches zero, at which point the wormhole vanishes. Additionally, we observe that the extracted MI does not always increase monotonically with the throat size; as the distance between the detectors and the wormhole throat increases, the influence of the wormhole diminishes substantially, although the overall trends in the harvested correlations remain consistent.
	\begin{figure}[b]
		\centering{\subfigure{\includegraphics[width=0.98\linewidth]{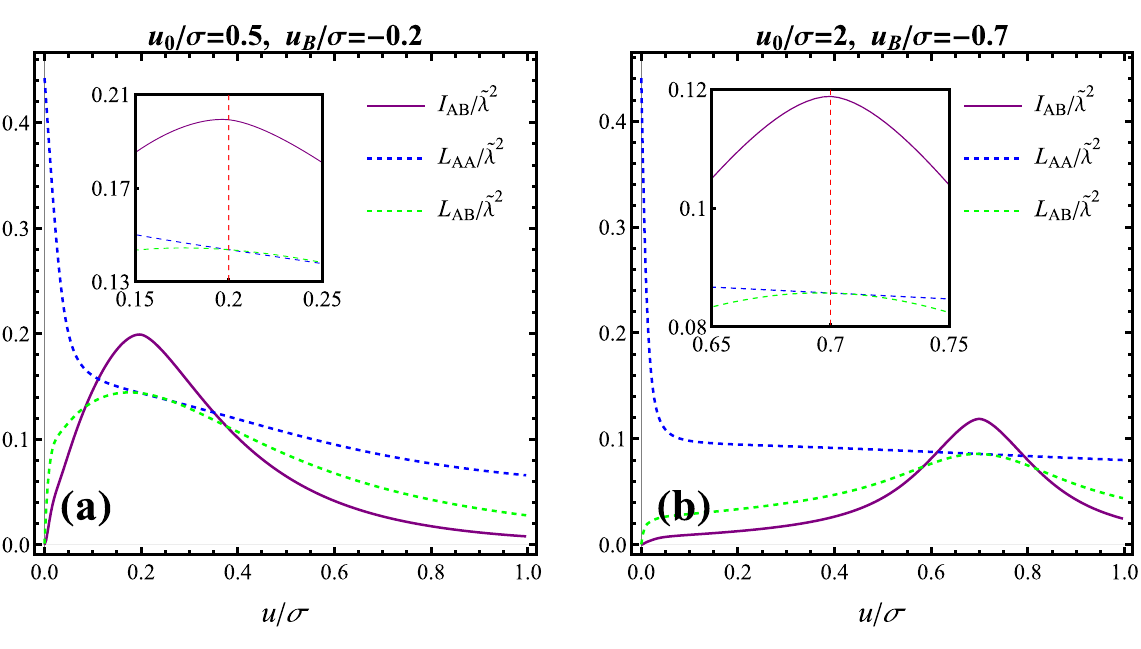}}
		}
		\caption{
			The plots illustrate the extracted MI (solid purple), TP (dashed blue), and the off-diagonal element \(L_{AB}\) (dashed green). In panel (a), the wormhole throat is set to 0.5 and detector Bob is fixed at \(u_0/\sigma = -0.2\), while in panel (b), the size is set to 2 and Bob is fixed at \(u_0/\sigma = -0.7\), with Alice positioned at various locations. The insets emphasize the regions of peak MI, with the red dashed line marking the position symmetrical to Bob. Parameters are set as \(\Omega \sigma = 1\) and \(M = 10^{-2}\).}
		\label{fig_5}
	\end{figure}
	
	The correlation harvesting process is influenced not only by parameters such as the wormhole throat size but also by the positions of the particle detectors, which exert a non-negligible effect. Therefore, to further elucidate these positional dependencies, we subsequently investigate the impact of the detector distribution while keeping the wormhole throat size fixed. Specifically, we first examine how the MI and TP vary with the symmetrically placed positions of the two detectors, denoted as \( u \). The results are presented in Fig.~\ref{fig_schBi}. Interestingly, neither MI nor TP exhibits a simple monotonic relationship with the detector separation. For small wormhole throats, as shown in Figs.~\ref{fig_schBi}(a) and (b), both MI and TP initially increase sharply with growing detector separation, rapidly attaining a peak value before gradually decaying. The peak height depends strongly on the throat size: as the throat size decreases, the peak becomes higher, and the MI declines rapidly with increasing distance \( u \) from the throat. For sufficiently small throats, the MI nearly drops to zero at larger separations. In contrast, for larger wormhole throats, as illustrated in Fig.~\ref{fig_schBi}(c), the peak structure vanishes, and both MI and TP decrease monotonically with increasing detector separation. In fact, larger throats yield higher MI values at greater distances from the throat compared to small throats, albeit at the expense of a reduced maximum MI. The disappearance of the MI peak stems from the rapid increase in MI near the wormhole throat as the throat size grows, coupled with a slower decay at the original peak positions. Consequently, when the near-throat MI surpasses the values at outer positions, the overall MI decreases monotonically with distance, as depicted in Fig.~\ref{fig_schBi}(c).
	
	Notably, the extracted MI does not vanish even at the wormhole throat itself and persists unless the wormhole vanishes entirely. For small throat sizes, the maximum MI occurs at a finite distance from the throat. However, as the throat size increases, this peak height diminishes progressively until it disappears altogether. For a sufficient larger throats, the harvested correlations are higher the closer the detectors are to the throat. In this context, the presence of the wormhole markedly enhances the harvesting of MI. During the initial phase of throat increasing, the presence of the wormhole leads to a rapid increase in MI, highlighting the strong correlation established between the two particle detectors. However, as the throat size continues to grow, although the peak in MI steadily diminishes, it facilitates the harvesting of MI by detectors at distances farther from the horizon.
	\begin{figure}[b]
		\centering{\subfigure{\includegraphics[width=0.98\linewidth]{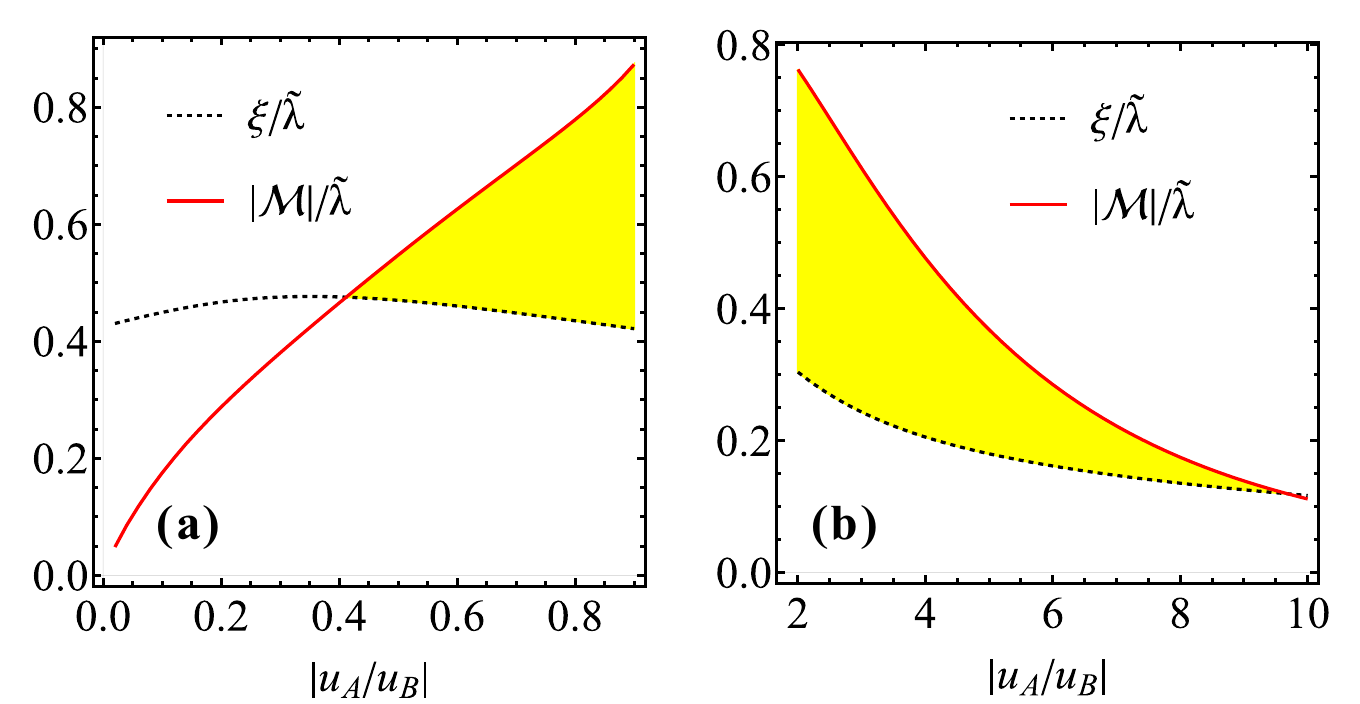}}
		}
		\caption{The plots illustrate the nonlocal contribution term \(\mathcal{M}\) (in red) and the noise term \(\xi\) (in black dashed line) for a wormhole throat set at \(u_0 / \sigma = 1\). Detector Bob is fixed at \(u_B / \sigma = -0.05\), while detector Alice is positioned at various locations. Panels (a) and (b) respectively depict the cases where Alice is farther from and closer to the wormhole throat than Bob. The yellow shaded region indicates the areas where \(\mathcal{M} > \xi\). The parameters are set as \(\Omega \sigma = 1\) and \(M = 10^{-2}\). }
		\label{fig_noise}
	\end{figure} 
	
	To delve deeper into the impact of detector positions on the harvested correlations, we build upon the aforementioned fixed wormhole throat size by anchoring detector Bob at a position \( u_B \) on the left side of the throat and continuously varying the position \( u_A \) of detector Alice, located on the right side of the wormhole, thereby investigating the effects of this asymmetric placement. Due to the asymmetric placement of detectors Alice and Bob, their TP differ, i.e., \( L_{AA} \neq L_{BB} \), until detector Alice reaches a symmetrical position relative to Bob. Furthermore, the off-diagonal element \( L_{AB} \) does not match the TP values. Fig. \ref{fig_5} illustrates the MI in solid purple, the TP in dashed blue, and \( L_{AB} \) in dashed green. As expected, the MI trend is primarily influenced by the off-diagonal term \( L_{AB} \). As detector Alice moves further from the wormhole throat, \( L_{AB} \) gradually increases, reaching a peak value equal to \( L_{AA} \) at the symmetrical position, and subsequently decreases. The MI also peaks at this position, indicating optimal MI at the symmetrical location relative to Bob, as shown in the subfigure.
	
	The wormhole size does not significantly impact the MI peak value but does affect the shape of the peak region. Notably, within this asymmetric configuration, the MI progressively diminishes as detector Alice approaches the wormhole throat. This can be attributed to the extreme relative Hawking temperature \cite{Bueley:2022ple}. Overall, this peak region suggests that the wormhole bridges two special image-symmetrical locations, characterized by an optimal correlation.
	\begin{figure}[t]
		\centering{\subfigure{\includegraphics[width=0.99\linewidth]{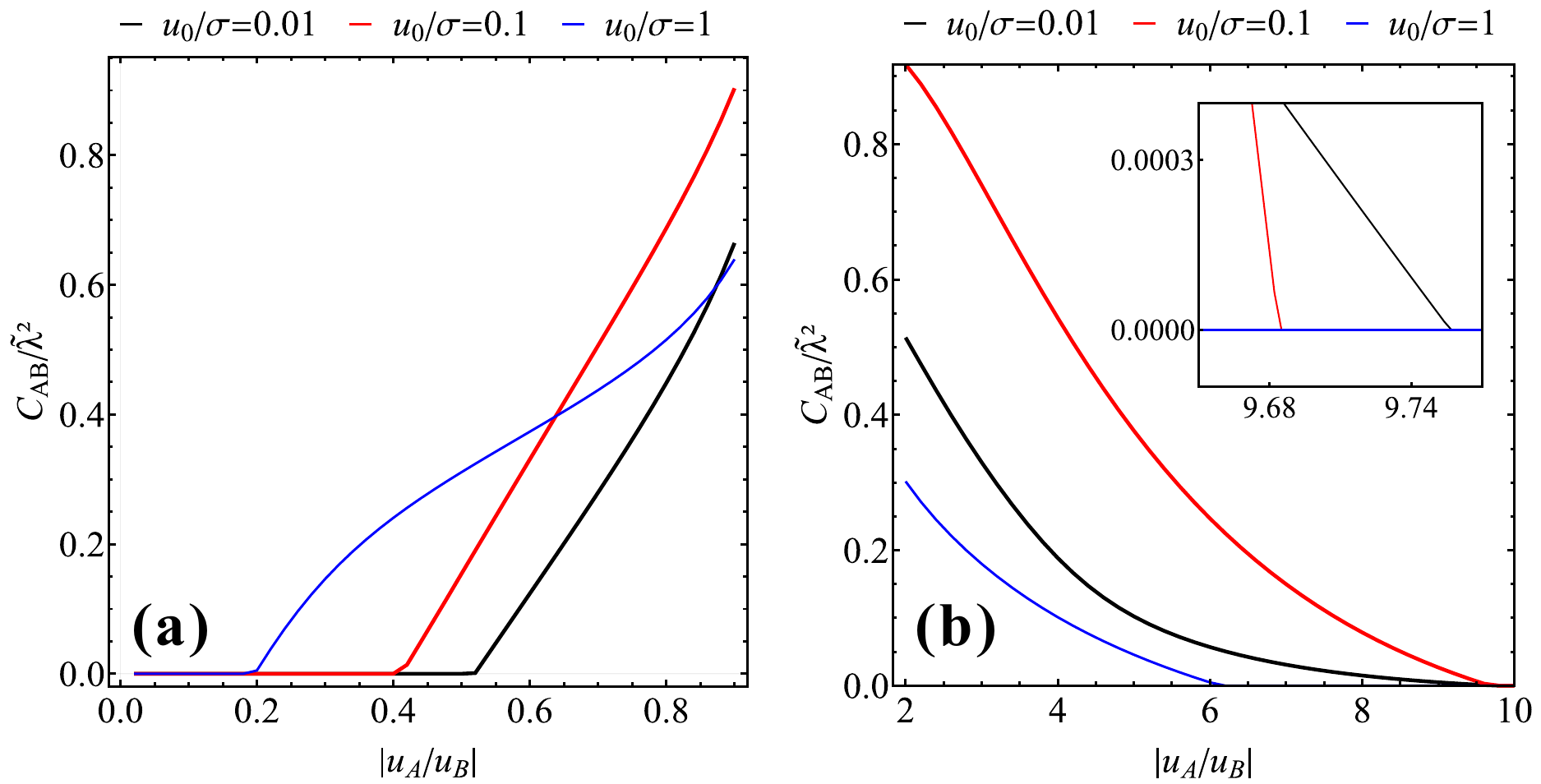}}
		}
		\caption{The plots illustrate the extracted concurrence for different wormhole throat sizes, with the throat set at \(u_0/\sigma = 0.01\) (black), \(u_0/\sigma = 0.1\) (red), and \(u_0/\sigma = 1\) (blue). Detector Bob is fixed at \(u_B/\sigma = -0.05\), while detector Alice is positioned at various locations. Panels (a) and (b) respectively depict the cases where Alice is farther from and closer to the wormhole throat than Bob. The parameters are set as \(\Omega \sigma = 1\) and \(M = 10^{-2}\).}
		\label{fig_CAB}
	\end{figure} 
	
	Finally, since the MI encompasses both classical and quantum correlations, we conduct a deeper analysis of the nonlocal correlation component within the harvested correlations by computing the nonlocal contribution term \(|\mathcal{M}|\) (in red) and the classical noise term \(\xi\) (in black dashed line), as shown in Fig.~\ref{fig_noise} (a) and (b). We find that, for a fixed wormhole throat size, the nonlocal contribution term \(|\mathcal{M}|\) achieves its maximum value near the mirror-symmetric position of detector Alice relative to Bob. The nonlocal contribution term \(\mathcal{M}\) (in red) is more sensitive to changes in the position of detector Alice, whereas the classical noise term \(\xi\) (in black dashed line) varies relatively slowly with respect to it. This leads to a rapid decline in \(|\mathcal{M}|\) as detector Alice moves away from the mirror-symmetric position relative to Bob (with the yellow shaded region in the figure denoting the parameter regime where \(|\mathcal{M}| > \xi\)); beyond a certain distance, \(|\mathcal{M}|\) falls below \(\xi\). This implies that the nonlocal correlations decay swiftly to zero far from the mirror-symmetric configuration. 
	
	For a more intuitive visualization of how these nonlocal correlations vary with the position of detector Alice, in Fig.~\ref{fig_CAB}(a) and (b) we further compute and plot the concurrence between the two detectors for three different wormhole throat sizes: \(u_0 / \sigma = 0.01\) (black), \(u_0 / \sigma = 0.1\) (red), and \(u_0 / \sigma = 1\) (blue). In all three cases, the concurrence reaches its maximum when the two detectors are at mirror-symmetric positions relative to the throat and decays rapidly upon deviation from this configuration, culminating in entanglement sudden death. This indicates that the correlations harvested by the detectors encompass both classical local correlations and nonlocal quantum correlations, particularly when the two detectors are positioned at mirror symmetry relative to the throat, where the nonlocal correlations achieve their maximum value.
	
\section{conclusion}\label{sec_IV}
	In summary, we employ the UDW detectors model to investigate how the size of the wormhole throat and the distribution of detectors influence the extracted correlation. Our results indicate that the presence of a wormhole, characterized by a nonzero throat size, enables detectors located on opposite sides to extract MI. Furthermore, when detector Bob is fixed, we observe a peak in the extracted MI, with the maximum occurring when Alice is positioned symmetrically. Notably, the extracted MI remains nonzero as long as the throat size is nonzero; although it diminishes rapidly as the distance between the detectors and the throat increases. These peaks suggest that the wormhole acts as a conduit between two specific symmetrical locations, resulting in the MI extracted between them reaching an optimal value. Subsequently, by analyzing the behavior of the nonlocal contribution term and the classical noise term, as well as by computing the concurrence between the two detectors, we demonstrate that the harvested correlations contain nonlocal correlations. These findings demonstrate that the existence of a null-like wormhole can enhance quantum correlation behaviors between classically causally disconnected regions. It is essential to emphasize, nevertheless, that such correlation harvesting phenomena do not enable inter-universe communication, since classical signals remain unable to traverse the wormhole.
	
	In the framework of quantum gravity theory, wormholes may no longer be just singular solutions in classical theories, but rather fundamental structural units of quantum spacetime, closely linked with quantum entanglement, information transfer, and spacetime topology. Future studies could compare the phenomenon of correlations harvesting in wormhole spacetimes across different theoretical frameworks, such as loop quantum gravity \cite{Brahma:2020eos,Lewandowski:2022zce,Zhang:2024khj,Zhang:2024ney,Liu:2024soc,Liu:2024iec}, string theory \cite{Kanti:2011jz}, and modified gravity \cite{Moffat:2014aja,Moffat:2013sja,Liu:2023uft,Liu:2024lbi}, to gain a deeper understanding of how these theoretical modifications affect the quantum properties of wormholes.
\section{Acknowledgments}
The authors gratefully acknowledge Shu-Min Wu for insightful discussions.
This work was supported by the National Natural Science Foundation of China under Grants  No. 12475051, No. 12375051, and No. 12421005; the science and technology innovation Program of Hunan Province under grant No. 2024RC1050; the Natural Science Foundation of Hunan Province under grant No. 2023JJ30384; and the innovative research group of Hunan Province under Grant No. 2024JJ1006.

\renewcommand{\appendixname}{Appendix}

\section{Appendix}\label{sec_V}
\appendix
\section{ Details of computing the element of RDM}
Throughout this study, we disregard angular effects by setting $\phi_A=\phi_B$. 
We  utilize the dimentionless coupling coefficient $\tilde{\lambda}:=\lambda \sqrt{\sigma}$. Our investigation focuses on how the correlation changes with variations in the size of the wormhole throat the detectors' positions.

Here, we provide detailed computations for the elements of the final RDM $\rho_{AB}$. According to equations (\ref{LAABB}-\ref{eq:noise}), to obtain the final extracted MI or concurrence, it is essential to compute the elements $\mathcal{L}_{AB},\mathcal{L}_{DD}$ and $|\mathcal{M}|$. By employing the contour integration method \cite{Bueley:2022ple}, these elements can be simplified to the following form:
\begin{widetext}
	\begin{align}
		\begin{split}
			\mathcal{M}=&-2 K_+ \sum_{n=-\infty}^{+\infty} \int_{0}^{\infty}\mathrm{d}z\; e^{-a_{AB}\, z^2}\mathrm{cos}(\beta^-_{AB}\, z) \left[\frac{1}{\sqrt{\mathrm{cosh}\,\alpha^-_{AB,n}-\mathrm{cosh}\, z}}-\frac{\xi}{\sqrt{\mathrm{cosh}\,\alpha^+_{AB,n}-\mathrm{cosh}\, z}}\right], \\
			\mathcal{L}_{AB}=&2 K_- \sum_{n=-\infty}^{+\infty} \lim_{R\to \infty} \mathrm{Re}\int_{0}^{R+i\eta}\mathrm{d}z e^{-a_{AB}\, z^2-i\beta_{AB}\, z} \left[\frac{1}{\sqrt{\mathrm{cosh}\,\alpha^-_{AB,n}-\mathrm{cosh}\, z}}-\frac{\xi}{\sqrt{\mathrm{cosh}\,\alpha^+_{AB,n}-\mathrm{cosh}\, z}}\right], \\
			\mathcal{L}_{DD}=&\frac{\lambda^2\sigma^2}{2}\int_{\mathbb{R}}\mathrm{d}x \frac{e^{-\sigma^2(x-\Omega)^2}}{e^{x/T_D}+1} -\xi \frac{\lambda^2\sigma}{2\sqrt{2\pi}} \mathrm{Re}\int_{0}^{\infty}\mathrm{d}x \frac{e^{-a_Dx^2-i\beta_Dx}}{\sqrt{\mathrm{cosh}\alpha_{D,0}^{+}-\mathrm{cosh}\,x}}\\
			&+\frac{\lambda^2\sigma}{\sqrt{2\pi}}\sum_{n=1}^{+\infty}\mathrm{Re}\int_{0}^{\infty}\mathrm{d}x\, e^{-a_Dx^2-i\beta_Dx} \left(\frac{1}{\sqrt{\mathrm{cosh}\,\alpha^-_{AB,n}-\mathrm{cosh}\, x}}-\frac{\xi}{\sqrt{\mathrm{cosh}\,\alpha^+_{AB,n}-\mathrm{cosh}\, x}}\right), 
		\end{split}
	\end{align}
	where
	\begin{align}
		&\alpha^\pm_{AB,n}:=\mathrm{arccosh}\left[\frac{u_0^2}{l^2\gamma_A\gamma_B}\left(\frac{(u_A^2+u_0)(u_B^2+u_0)}{u_0^2}\mathrm{cosh}\left[2\pi n u_0/l\right]\pm 1\right)\right], \\
		&a_{AB}:=\frac{\gamma_A^2\gamma_B^2l^4}{2\sigma^2(\gamma_A^2+\gamma_B^2)u_0^2}, \\
		&\beta^{\pm}_{AB}:=\frac{\gamma_A\gamma_B(\gamma_A \pm \gamma_B) l^2 \Omega}{(\gamma_A^2+\gamma_B^2)u_0},\\
		&K_{\pm}:=\frac{\lambda^2\sigma\sqrt{\gamma_A\gamma_B}}{4\sqrt{\pi(\gamma_A^2+\gamma_B^2)}}exp\left[-\frac{\Omega^2\sigma^2(\gamma_A \pm\gamma_B)^2}{2(\gamma_A^2+\gamma_B^2)}\right],\\
		&\gamma_D= \frac{\sqrt{u^2(u^2+2u_0)}}{l},\; T_D=u_0/(2\pi l^2 \gamma_D),\\
		&\alpha^\pm_{D,n}=\alpha^\pm_{DD,n},\quad a_{D}=a_{DD},\quad \beta_{D}=\beta_{DD},\;\qquad D\in\{A,B\}.
	\end{align}
	
\end{widetext}

%


\end{document}